# Investor Sentiment and Market Movements: A Granger Causality Perspective

Tamoghna Mukherjee

Department of Computer Science & Engineering,
Amity School of Engineering & Technology
Amity University Kolkata, India
Email: tamoghna.9@hotmail.com
ORCID ID: 0000-0001-8065-3850

**Abstract**

The stock market is heavily influenced by investor sentiment, which can drive buying or selling behavior. Sentiment analysis helps in gauging the overall sentiment of market participants towards a particular stock or the market. Positive sentiment often leads to increased buying activity and vice versa. Granger causality can be applied to ascertain whether changes in sentiment precede changes in stock prices. The study is focused on this aspect and tries to understand the relationship between close price index and sentiment score with the help of Granger causality inference. The study finds a positive response through hypothesis testing.

Keywords: Stock Market, Investor Sentiment, Sentiment Analysis, Granger Causality, Hypothesis Testing

**Introduction**

Relationship between close price index of Stock Market and sentiment score is an interesting unexplored aspect in terms of sentiment analysis paradigm on stock index. In this study we try to prove such a relationship exists by using Granger's causality inference. To understand the idea of sentiment score and close price we highlight the idea of daily transactions in the stock exchange.

The closing price of a security refers to its final transaction price before the market formally closes to normal trading. Investors occasionally use it as a benchmark to determine how well a company has performed over the last day. A statistic for assessing client sentiment is sentiment scores. Even though there are numerous aspects that affect how a stock exchange functions, psychological factors like user perceptions of regulatory changes, new investments, or natural catastrophes can have a significant impact. The market's assessment of the firm, customer perception of the brand, and reasons other than just financial ones all have an impact on stock price changes. Thus, investor mood, industry reports, social media reviews, and media attitude about any firm may all offer valuable insights into the dynamics of a stock exchange's opening and closing price.

Based on this understanding we tried to state that for an analytical assertion attempt to detect if a certain trend is influential and present helpful results to predict a different trend is the Granger causality test [6]. To check how the closing price Y for today depends on the sentiment score X of the same day, our understanding would be that Granger states that the value of Y depends on the value of X. If this is untrue, we argue that the X does not directly affect Y. In this context, our study uses the Granger causality test to determine if the Stock Exchange Close Price and Sentiment Score are related to each other and how the later affects the value of the former.

**Literature Review**

Financial news has a significant impact on financial markets as a whole. One of the primary motivations for the search for the sentiment score from news headlines was discovering a relationship between regular transactions in Tradingmarkets and people's sentiments. A suggestion that was made to support the assertion suggested that daily net sentiment and daily stock closing prices were positively correlated [1]. A machine learning approach that uses previous stock price data and financial news to forecast future stock prices is suggested in a further insight [2]. The blending of diverse feelings was considered based on market data and sentiment indexes from different news sources. Then, in order to anticipate the movement of the stock market index, we built a number of basic classifiers using the recurrent neural network and combined them using the evidential reasoning rule [3]. A recommendation for investors and market authorities on risk management of the green stock market is provided by the empirical study's findings, which show the sophisticated risk contagion mechanism in the green stock market [4]. The study that investigates the research finding that ESG news sentiment helps to reduce stock price crash risk and extends the research on the governance mechanism of stock price crash risk emphasizes the risk factor based on sentiment trend [5].

To properly understand Granger causality, it was intended in this context to understand the link between quotes from the food industry enterprises and the major market indicators [6]. It was found that the long- and short-term aspects of this connection had not been examined in the context of various shocks when analyzing the causal link between gold and stocks [7]. The paper examined dynamic causal connections between the prices of precious metals and US stock market indices over all quantiles of the conditional distribution within the same framework [8]. The study continues the reference by examining the Granger causation between US-China political ties and Chinese stock market returns using a rolling-window technique and the bootstrapping method [9]. The research that demonstrated nonlinear causation between the US stock market and the Greek and Spanish stock markets, Granger, also placed emphasis on the interdependence of the markets [10]. We conducted our study using the aforementioned line of reasoning, and for five years, we applied the information in Bombay Stock Exchange (BSE) intraday market transactions [11].

**Methodology**

We have conducted the analysis in two folds where in the first instance we have calculated the sentiment score of daily news headlines related to the Indian financial domain for a period of five years starting from May, 2015 to May, 2020.

In the second part we have applied the Granger causality hypothesis testing between Closing Price of BSE for the same five year period and sentiment score obtained from the first phase. In the following subsections both the phases are explained in terms of algorithmic approach.

1. **Calculation of Sentiment Score**
   A confidence score was derived that spans from 0 to 1, with 1 denoting extreme confidence and 0 denoting extreme uncertainty, in order to calculate the emotion score. The process is as follows:
   Step 1: Preparing text to word with Flair
   Step 2: Word Embeddings with Flair
   Step 3: Vectorizing the text
   Step 4: Partitioning the data for Train and Test Sets
   Step 5: Obtaining the scores
2. **Granger's Causality Test**
   The test was used to determine whether variance in sentiment score influences the Close Price of Stock Market Exchange. The detailed methodology are as follows:
   Step 1: We verify the time series data to make sure there are enough observations for a trustworthy analysis.
   Step 2: Defining the null and alternative hypotheses where the alternative hypothesis contends that sentiment score does, in fact, affect close price, whereas the null hypothesis contends that sentiment does not.
   Step 3: Calculation of the lag order.
   Step 4: Creation of a training set and a testing set out of the data. In this instance, the split is 80/20.
   Step 5: Testing the hypotheses.
   Step 6: Analyzing the outcomes

**Observation**

Sentiment Score obtained from intraday news headlines was taken into consideration with Close Price of BSE to check the existence of Granger Causality between the two.

**Part 1:**

Proposed null hypothesis - Sentiment score does not influence close price.
Proposed alternate hypothesis - Sentiment score influences close price.

|       | p value |
|-------|---------|
| Lag 1 | 0.0184  |
| Lag 2 | 0.0074  |
| Lag 3 | 0.0156  |
| Lag 4 | 0.0384  |

In this case since p value is low, hence null hypothesis is rejected. Alternate hypothesis is accepted.

**Part 2:** The Granger causality test repeated in the other manner.

Proposed null hypothesis - Close index does not influence sentiment score.
Proposed alternate hypothesis - Close index influences sentiment score.

|       | p value |
|-------|---------|
| Lag 1 | 0.1737  |
| Lag 2 | 0.1810  |
| Lag 3 | 0.3321  |
| Lag 4 | 0.3452  |

In this case since p value is high, hence null hypothesis is accepted. Alternate hypothesis is rejected.
Hence we can state that Sentiment Score influences Close Price.

**Conclusion**

The study concludes that by employing Granger causality tests, analysts can assess whether sentiment data has predictive power over stock prices. If sentiment changes consistently precede price movements, it suggests that sentiment analysis can be a valuable tool for forecasting stock price trends.